\documentstyle{aipproc}

\begin{document}
\title{``Signature'' neutrinos from photon sources at high redshift}

\author{Marieke Postma}
\address{Department of Physics and Astronomy, UCLA, Los Angeles, CA
90095-154
\\E-mail: postma@physics.ucla.edu}

\maketitle

\begin{abstract}
The temperature of the cosmic microwave background increases with
redshift; at sufficiently high redshift it becomes possible for
ultrahigh-energy photons and electrons to produce muons and pions
through interactions with background photons. At the same time, energy
losses due to interactions with radio background and intergalactic
magnetic fields are negligible. The energetic muons and pions
decay, yielding a flux of ``signature'' neutrinos with energies $E_\nu
\sim 10^{17}$eV. Detection of these neutrinos can help understand the
origin of ultrahigh-energy cosmic rays.
\end{abstract}

The origin of ultrahigh-energy cosmic rays~\cite{mp:data} with
energies beyond the Greisen-Zatsepin-Kuzmin (GZK) cutoff~\cite{mp:gzk}
remains an outstanding puzzle~\cite{mp:reviews}. Most of the proposed
explanations can be categorized in two main classes.  In the {\em
bottom-up/acceleration} scenarios, charged particles are accelerated to
ultrahigh energies in giant astrophysical ``accelerators'', such as
active galactic nuclei and radio galaxies~\cite{mp:conv,mp:mpr}.  In
the {\em top-down/decay} scenario on the other hand, massive objects
such as topological defects
(TD)~\cite{mp:TD_review,mp:TD,mp:TD_ber,mp:TD_nu} and superheavy relic
particles~\cite{mp:particles,mp:gk2,mp:kt} decay, emitting mainly
ultrahigh-energy (UHE) photons.  TD and relic particles can exist at
much higher redshift than the astrophysical candidate sources, which
are formed only after the onset of galaxy formation. To understand the
origin of the ultrahigh-energy cosmic rays (UHECR), it is crucial to
distinguish between these two different scenarios.  In this talk we
will show that the observation of a diffuse flux of ``signature''
neutrinos with energies $E_\nu \sim 10^{17}$eV~\cite{mp:kp} can
provide evidence for UHE photon sources at high redshift --- for
members of the {\em top-down/decay} class.

At small redshift UHE photons lose their energy mainly through
scattering off the radio background (RB), and in the subsequent
electromagnetic cascade~\cite{mp:berezinsky,mp:reviews}. In addition,
the cascade electrons lose energy through synchrotron radiation in the
intergalactic magnetic field (IGMF). This picture changes completely
at high redshift, where both radio background and magnetic fields are
small. At redshift $z$ the density of CMB photons is higher by a
factor $(1+z)^3$, while the density of radio background is either
constant or, more likely, lower. At some redshift $z_{_R}$ the
scattering of high-energy photons off CMBR dominates over their
scattering off RB.  Based on the analyses of
refs.~\cite{mp:condon,mp:recent} we take $z_{_R} \sim 5$. Also, before
galaxy formation the IGMF is weak, and for $z> z_{_M}\sim 5$
synchrotron losses are not significant.

But the absence of RB and IGMF at early times is not the only
difference: at redshift $z$ the temperature of the cosmic microwave
background radiation increases by a factor $(1+z)$. Because of this,
at high redshift the production of muons and pions in interactions of
UHE photons and electrons with the CMBR becomes possible.  Muons and
charged pions decay into neutrinos, which can reach us today. The
threshold for muon production is $\sqrt{s}> 2 m_\mu=0.21$GeV, or
\begin{equation} 
E_{\gamma,e} > E_{\rm th}(z)=\frac{ 10^{20}{\rm eV}}{1+z} 
\label{mp:threshold} 
\end{equation} 
 
But will muons, and thus neutrinos, indeed be produced? To answer this
question one has to look closely at the propagation of UHE photons at
$z \stackrel{>}{_{\scriptstyle \sim}} 5$. Photons scatter off  CMB
photons and, if their energy is above the threshold for muon pair
production,  given in~eq.~(\ref{mp:threshold}), they can either produce a
muon pair, an electron pair or a double electron pair.  Among these
processes electron pair production (PP) has the highest cross section
for photon energies $E_\gamma \stackrel{<}{_{\scriptstyle \sim}} 5
\times 10^{20} {\rm eV} /(1+z)$. Since the energies of the two
interacting photons are vastly different, either the electron or the
positron from PP has energy close to that of the initial photon.  At
higher photon energies, double pair production (DPP) becomes more
important~\cite{mp:dpp}. Four electrons, each carrying about $1/4$ of
the initial photon energy, are produced in this reaction.  Thus, after
an initial $\gamma \gamma_{_{CMB}}$ interaction one ends up with one
or more UHE electrons.

These electrons in turn scatter off CMBR. For electron energies above
the muon threshold, inverse Compton scattering is negligible compared
to higher order processes~\cite{mp:reviews}, such as triplet
production (TPP) $e \gamma_{_{CMB}}\rightarrow e\, e^+ e^-$ and muon
electron-pair production (MPP) $e \gamma_{_{CMB}}\rightarrow e\, \mu^+
\mu^-$.  For $\sqrt{s}> 2m_{\pi^{\pm}} = 0.28$GeV charged pion
production may also occur through $e \gamma_{_{CMB}}\rightarrow e\,
\pi^+ \pi^-$.  The TPP cross section is larger than that for MPP
\cite{mp:tpp,mp:tpp2}. However, for center of mass energies $s \gg
m_e^2$ the inelasticity for TPP is very small: one of the 
electrons produced in TPP carries $99.9\%$ of the incoming electron's
energy.  It can interact once again with the CMBR.  As a result, the
leading electron can scatter many times before losing a considerable
amount of energy; and with every scattering there is another chance to
produce a muon pair. So instead of comparing cross sections, one
should compare the {\em energy attenuation} length for TPP with the
{\em interaction} length for MPP.  As the latter is much larger, all
electrons with energies above muon threshold will produce muons (or
pions).

Topological defects and relic particles emit UHE photons.  One can
parameterize the time-dependence of their photon production rate as
$\dot{n}_{_\gamma} =\dot{n}_{_\gamma,0} (t/t_0)^{-m}$, with $m=0$ for
decaying relic particles, $m=3$ for ordinary strings, monopolonium and
necklaces, and $m \ge 4$ for superconducting
strings~\cite{mp:TD_review,mp:TD,mp:TD_ber}.  Integrating
$\dot{n}_{_\gamma} =\dot{n}_{_\gamma,0}$ over time, taking redshift
into account, yields a neutrino flux:
\begin{eqnarray}  
n_\nu & = & \xi \int^{z_{\rm max}}_{z_{\rm min}} dt \
\dot{n}_{\gamma}(z) \ (1+z)^{-4} \nonumber \\ & = & \xi \frac{3}{-2a}
\dot{n}_{\gamma,0} \, t_0 \, [(1+z_{\rm min})^a-(1+z_{\rm max})^a] .
\label{mp:nu_z}  
\end{eqnarray}  
Here $z_{\rm min}=\max (z_{_R},z_{_M}) \approx 5$ the minimum redshift
at which both RB and IMGF are negligible, $z_{\rm max} \sim 3 \times
10^3$ the redshift at which the universe becomes transparent to UHE
neutrinos~\cite{mp:ggs}, and $\xi$ is the number of neutrinos produced
per UHE photon.  We take $\xi \approx 4$, as one UHE photon produces
one UHE electron, which generates a pair of muons whose decay gives
four neutrinos.  This is probably an underestimate because DPP
produces more than one UHE electron.  Also, the electron produced
alongside the muon pair in MPP may have enough energy for a second
round of muon pair-production.

For $m<11/3$, $a< 0$, and, according to~eq.~(\ref{mp:nu_z}), most of
neutrinos come from red shift $z\sim z_{\rm min}\approx 5$. All these
neutrinos are produced by photons with energies $E_\gamma > E_{\rm
min} = 10^{20}{\rm eV}/(1+z_{\rm min}) \sim 2 \times 10^{19}$eV.  If
decaying TD or relic particles are the origin of the UHECR today,
one can use the observed UHECR flux to fix the overall normalization
constant $\dot{n}_{\gamma,0}(E>E_{\rm min})$.  Using the photon fluxes
calculated in~\cite{mp:TD_ber,mp:pj,mp:ps}, we get for the neutrino
flux:
\begin{equation} 
\phi_{\nu} \sim \left \{ 
\begin{array}{lll} 
 10^{-21}{\rm cm}^{-2} {\rm s}^{-1} {\rm 
  sr}^{-1}, & \quad {\rm relic \: particles}\, (m=0), \\  
 10^{-18}{\rm cm}^{-2} {\rm 
  s}^{-1} {\rm sr}^{-1},  & \quad {\rm monopolonium} \, (m=3) , \\   
10^{-16}{\rm cm}^{-2} {\rm s}^{-1} {\rm 
  sr}^{-1}, &\quad {\rm necklaces}\, (m=3). \\ 
\end{array} \right.  
\label{mp:flux} 
\end{equation} 
The difference in flux between monopolonium and necklaces lies in
their different clustering properties: whereas monopolonium (and relic
particles) clusters in galaxies, necklaces are uniformly distributed
throughout the universe.

The energy of these neutrinos can be estimated by
\begin{equation}
  E_\nu \sim \frac{10^{20} {\rm eV}}{1+z} \times \frac{1}{4} \times
         \frac{1}{3} \times \frac{1}{1+z}.
\end{equation}
As the photon spectrum of TD is a sharply falling function of
energy, most neutrinos will come from photons with energies just
above the threshold.  In MPP, each muon produced gets about $1/4$ of the
incoming electron energy~\cite{mp:tpp2}. A muon decays into an
electron and two neutrinos, each getting approximately $1/3$ of the
muon energy.  Furthermore, the neutrino energy is redshifted by a
factor $1/(1+z)$.  As already noted before, most neutrinos come from
$z \sim z_{\rm min} \approx 5$; their energy is $E_\nu \sim
10^{17}$eV.

Neutrinos from slowly decaying relic particles or some other $m=0$
source are too sparse to be detectable in the foreseeable
future. However, the flux from $m=3$ sources, especially from
necklaces, may be detected soon. This flux exceeds the background flux
from the atmosphere and from pion photo-production on CMBR at this
energy~\cite{mp:hs,mp:stecker,mp:sp}, as well as the fluxes predicted
by a number of models~\cite{mp:whitepaper}. TD can produce a large
flux of primary neutrinos.  However, the primary flux peaks at
$E_\nu\sim 10^{20}$eV, while the secondary flux peaks at $E_\nu\sim
10^{17}$eV and creates a distinctive ``bump'' in the spectrum.  Models
of active galactic nuclei (AGN) have predicted a similar flux of
neutrinos at these energies~\cite{mp:mpr}.  The predictions of these
models have been a subject of debate~\cite{mp:wb}.  However, every one
agrees that AGN cannot produce neutrinos with energies of $10^{20}$eV.
So, an observation of $10^{17}$eV neutrinos accompanied by a
comparable flux of $10^{20}$eV neutrinos would be a signature of a TD
rather than an AGN.
 
Rapidly evolving $m \ge 4$ sources, {\em e.g.} superconducting
strings, have been ruled out as the origin of UHECR by the EGRET
bounds on the flux of low-energy $\gamma$-rays~\cite{mp:reviews}.
Although their density today is constrained, in the early universe these
sources might have existed in large enough numbers to produce a
detectable flux of neutrinos. Neutrinos with $E_\nu \sim 10^{17}$eV
are probably the only observable signature of such rapidly evolving
sources that were active at high red shift but are ``burned out'' by
now.
 
To conclude, we have shown that sources of ultrahigh-energy photons
that operate at red shift $z \stackrel{>}{_{\scriptstyle \sim}}5$
produce neutrinos with energy $E_\nu \sim 10^{17}$eV.  The flux
depends on the evolution index $m$ of the source.  A distinctive
characteristic of this type of neutrino background is a cutoff below
$10^{17}$eV due to the universal radio background at $z<z_{\rm
min}$. Detection of these neutrinos can help understand the origin of
ultrahigh-energy cosmic rays.

This work was supported in part by the US 
Department of Energy grant DE-FG03-91ER40662, Task C, and by a grant 
from UCLA Council on Research.


\end{document}